\documentclass[]{aastex7}

\usepackage{subfig}
\usepackage{rotating} 
\usepackage{array}      
\usepackage{longtable}  
\usepackage{appendix}  
\usepackage{soul} 
\usepackage{xcolor} 
\usepackage[most]{tcolorbox} 
\usepackage{booktabs}
\usepackage{threeparttable}
\usepackage{soul}      
\usepackage{xcolor}     
\sethlcolor{yellow}

\begin{document}

\title{Super-Jeans Fragmentation and Supply-Limited Accretion: Environment-Dependent Co-Evolution of Low- and High-Mass Cores}

\correspondingauthor{Guang-Xing Li}
\email{gxli@ynu.edu.cn}

\author[0000-0001-9798-9852]{Dan Miao}
\affiliation{Department of Astronomy, Xiamen University, Zengcuo’an West Road, Xiamen, 361005, China}
\email{miaodan@stu.xmu.edu.cn}

\author[0000-0003-3144-1952]{Guang-Xing Li}
\affiliation{South-Western Institute for Astronomy Research, Yunnan University, Chenggong District, Kunming 650500, China}
\email{gxli@ynu.edu.cn}

\author[0000-0002-7125-7685]{Patricio Sanhueza}
\affiliation{Department of Astronomy, School of Science, The University of Tokyo, 7-3-1 Hongo, Bunkyo, Tokyo 113-0033, Japan}
\email{patosanhueza@gmail.com}

\begin{abstract}

Protostellar core formation and growth in high-mass star-forming regions remain key to understanding massive star birth. 
We analyze the masses of 839 cores (resolved at scales of a few thousand au) from the ASHES project targeting 39 massive infrared dark cloud clumps.
The masses of the three most massive cores scale linearly with the total core mass. 
They maintain a constant mass fraction of $\sim 25\%$, $16\%$, and $10\%$ along the mass growth sequence. 
These fractions reveal that the progenitor seeds destined to become high-mass cores establish their mass dominance very early.
Additionally, the Gini coefficient (a statistical measure of inequality) of the core mass distributions increases along the mass growth sequence, confirming that the relative population of low-mass cores builds up toward later stages. 
This points to an environment-dependent fragmentation picture: central prestellar seeds rapidly evolve into high-mass cores via transport-driven super-Jeans fragmentation under rapid, non-stationary mass accumulation in high-density, turbulent hubs, subsequently maintain their dominance through supply-limited synchronized growth (at R$<1$~pc, $n_{\rm H_2} > 10^5~{\rm cm}^{-3}$), while the formation of the surrounding low-mass cores is relatively delayed due to their lower gas densities and the lack of non-stationary inflow acceleration effect, resulting in their continuous emergence through Jeans-like fragmentation in lower-density envelopes.
This picture is consistent with a gravity-driven scenario where the local free-fall time is the controlling factor. 
Our analysis suggests that non-stationary density-regulated fragmentation and supply-limited accretion jointly drive the synchronized co-evolution of the core cluster, seamlessly linking small-scale core growth with large-scale reservoir regulation.

\end{abstract}

\keywords{\uat{Star formation}{1569}--- \uat{Infrared dark clouds}{787} --- \uat{Protostars}{1302} --- \uat{Stellar accretion}{1578} --- \uat{Initial mass function}{796}}

\section{Introduction} 

Most stars form when giant molecular clouds (GMCs) collapse, typically producing groups that range from dozens to millions of members \citep[e.g.,][]{2003ARA&A..41...57L,2010ARA&A..48..431P,2020SSRv..216...64K}.
Recent kinematic studies indicate that these star-forming molecular clouds are undergoing global, multi-scale, and multi-epoch gravitational contraction \citep[e.g.,][]{2014A&A...561A..83P, 2018ApJ...852...12Y, 2019MNRAS.486..283B, 2025ApJ...979..233M}.
During this gravoturbulent collapse, fragmentation events give rise to dense clumps that serve as the primary sites for clustered star formation \citep{2004RvMP...76..125M,2007ARA&A..45..565M,2016SAAS...43...85K}.
A feature across these star-forming regions is hierarchical mass segregation. 
At large-scale scales, the most massive members concentrate in dense central hubs, while low-mass stars preferentially form along extended filamentary arms \citep[e.g.,][]{1997AJ....113.1733H, 1998MNRAS.295..691B, 2013ApJ...764...73P, 2018A&A...615A...9P, 2020A&A...642A..87K}.
Similarly, at the clump scale, relatively evolved protoclusters show robust evidence of mass segregation, with massive cores consistently located at the geometric center \citep[e.g.,][]{2024ApJS..270....9X}.
However, observations of pristine, early-stage infrared dark clouds (IRDCs) reveal a more complex reality: young clumps are subclustered and lack a well-defined geometric center. In this early regime, global geometric segregation is replaced by a profound density segregation \citep[e.g.,][]{2019ApJ...886..102S, 2023ApJ...950..148M}. 
This indicates that early fragmentation is governed by local physical conditions, giving rise to an environment-dependent process where progenitor seeds of massive cores preferentially form at localized density peaks, while lower-mass cores form in the more diffuse surrounding envelopes \citep[e.g.,][]{2023ApJ...950..148M, 2024ApJ...966..171M}.

Scenarios including classical turbulent fragmentation \citep[][]{2002ApJ...576..870P, 2004RvMP...76..125M}, turbulent core accretion \citep{2003ApJ...585..850M}, competitive accretion \citep{2001MNRAS.323..785B}, global hierarchical collapse \citep[GHC;][]{2019MNRAS.490.3061V}, and super-Jeans fragmentation —a recent model proposing that conditions such as rapid inflows or strong turbulence elevate the effective Jeans mass to directly form massive fragments \citep{2024MNRAS.528.7333L}— have been proposed to set the distribution of stellar masses and result in a mass-segregated cluster.
In classical turbulent fragmentation, supersonic turbulence fragments molecular clouds into a random network of dense cores, generating initial mass distributions primarily through shock compressions \citep{2002ApJ...576..870P, 2004RvMP...76..125M}.
As an extension of this framework, the later inertial-inflow model \citep{2020ApJ...900...82P} posits that while initial low-mass seeds are randomly generated by turbulence, massive stars require continuous mass supply from sustained, large-scale inertial inflows.
In the turbulent core accretion model, low-mass cores are supported primarily by thermal pressure, while massive stars form in regions where supersonic turbulence provides additional support against gravity, allowing a massive core to assemble before global collapse \citep{2003ApJ...585..850M}.
In competitive accretion, massive stars do not form from massive prestellar cores, but rather grow from low-mass seeds that outcompete by occupying the deepest parts of the cluster's global gravitational potential with higher accretion rates \citep{2001MNRAS.323..785B,2001MNRAS.324..573B}.
The GHC scenario posits a multi-scale, multi-epoch hierarchical contraction of molecular clouds: local high-density fluctuations have the shortest free-fall times and collapse rapidly into low-mass cores, while the global, large-scale cloud contraction proceeds more slowly, continually driving gas toward the central hubs. While fully consistent with competitive accretion at the local core scale, GHC emphasizes the continuous evolution of the parental cloud and global mass replenishment over several megayears \citep{2019MNRAS.490.3061V,2026MNRAS.547f2059V}.
Dense clumps can form quickly through converging flows, turbulence, or cloud collisions. 
When non-stationary, externally driven inflows cause the background density to increase on a mass-accumulation timescale, $t_{\rm acc}\equiv\rho/\dot{\rho}$, that is comparable to or shorter than the local free-fall time, $t_{\rm ff}\sim(G\rho)^{-1/2}$, transport-driven super-Jeans fragmentation can become important. 
Here, $t_{\rm acc}$ quantifies the density increase caused by external mass transport, whereas $t_{\rm ff}$ characterizes local collapse under the gas's own gravity. 
The background density changes substantially before small-scale perturbations can complete their self-gravitational collapse, so continuous mass injection favors larger fragmentation modes and increases the characteristic fragmentation mass \citep{2024MNRAS.528.7333L,10.1093/mnras/staf1116}.

The fundamental question then follows: which of these scenarios governs the emergence and evolution of cores of different masses$?$
First, classical turbulent fragmentation \citep{2002ApJ...576..870P, 2004RvMP...76..125M} posits that cores of varying masses, including high-mass ones, are generated almost simultaneously in the early stages via random supersonic turbulent compressions. 
Second, evolutionary models such as GHC \citep{2019MNRAS.490.3061V} and the inertial-inflow model \citep{2020ApJ...900...82P} predict that low-mass cores form first, with high-mass cores emerging later through prolonged competitive accretion or sustained large-scale inflows that build up their masses from low-mass seeds.
Conversely, the non-stationary, transport-driven super-Jeans fragmentation \citep{2024MNRAS.528.7333L,2024MNRAS.528L..52L} argues that converging inflows at the central hub enable the progenitor seeds to rapidly evolve into high-mass cores and establish mass dominance early. Meanwhile, the lower-density outer regions lack this non-stationary acceleration; their longer free-fall timescales naturally delay local fragmentation, resulting in the continuous, later emergence of low-mass cores.
Thus, by investigating whether the overall core population emerges simultaneously or sequentially, and specifically whether the progenitor seeds destined to become high-mass cores establish their mass dominance early or late, we can distinguish between these theoretical frameworks.
In star formation theory, a core's mass is set by its initial gravitational collapse mass and subsequent accretion. 
Indeed, this theoretical picture is supported by recent observational results from the ASHES and ALMAGAL surveys, which show that core populations have statistically higher masses in more evolved host clumps \citep{2024ApJ...966..171M, 2025A&A...696A.151C}. 
We can consider the core mass to increase over time through the accretion process. 
Therefore, we can use the ratio of total core masses to clump mass (the core formation efficiency) across the mass growth sequence as an indicator to study the fragmentation and growth issues mentioned earlier.

\section{Data and Results}

We use the mass of the 839 cores embedded in 39 clumps derived from the ALMA Survey of 70 $\mu$m dark High-mass clumps in Early Stages (ASHES; 2015.1.01539.S, 2017.1.00716.S, 2018.1.00192.S; PI: P. Sanhueza; \citealt{2019ApJ...886..102S,2023ApJ...950..148M}) project.

The ASHES project targeted a sample of 39 massive (220--4900 $M_{\odot}$), dense ($> 10^4\ \rm cm^{-3}$), and cold ($\sim 10$--20 K) 70 $\mu$m dark clumps located at distances between 2 and 6 kpc. 
These sources were selected as massive prestellar clump candidates lacking bright infrared emission to investigate the earliest phases of high-mass star formation, making them ideal candidates to be in the prestellar phase \citep{2023ApJ...950..148M}.
Using 1.3 mm mosaic observations from the ALMA 12 m and 7 m arrays, this project resolved 839 cores at spatial scales of a few $10^3$ au. 
The core masses are adopted from Table 4 of \citet{2023ApJ...950..148M}. 
Each clump contains between 8 and 39 cores, with an average of 21 cores per clump. 
The mass of the most massive core within each clump ranges from 1.7 to 81.1 $M_{\odot}$, with a mean value of 12.5 $M_{\odot}$. 
On clump scales (spatial resolution $\sim 10^5$ au), we adopt the refined clump masses from Table 1 of \citet{2023ApJ...950..148M}, which were derived by scaling the initial values using the 870 $\mu$m integrated fluxes obtained from Gaussian fitting.
Beyond the continuum-based characterization of the ASHES sample 
\citep{2019ApJ...886..102S,2023ApJ...950..148M,2024ApJ...966..171M,2026ApJ...997..155M}, several follow-up studies have investigated its outflow activity, chemistry, and dense-core dynamics \citep{2020ApJ...903..119L,2021ApJ...913..131T,2021ApJ...923..147M,2022ApJ...925..144S,2022ApJ...936...80S,2022ApJ...939..102L,2023ApJ...949..109L,2024ApJ...963..163I,2025ApJ...990..229L}, as well as the infall properties of ASHES cores \citep{2026ApJ...997..155M,2026ApJ..1004..207M}.

To investigate the mass relationship between individual cores and their host clumps, we plot the aforementioned data in Figure \ref{fig1}. We find that:

\begin{figure}[ht!]
\centering
\includegraphics[width=1\textwidth]{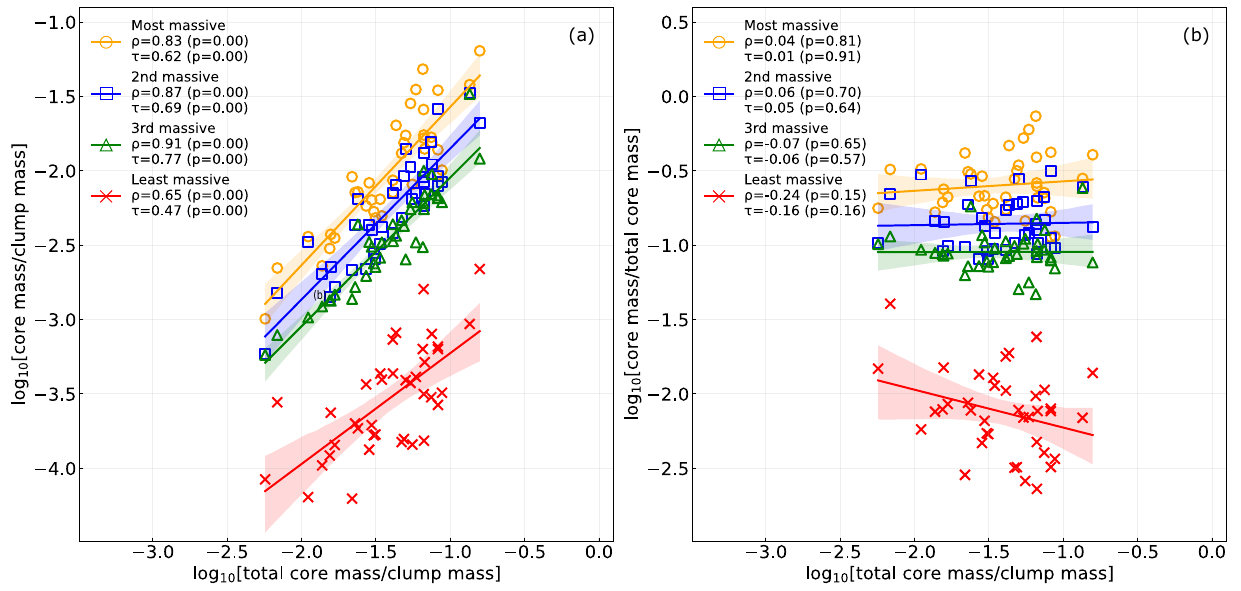}
\caption{Panel (a) shows the core-to-clump mass ratio ($M_{\rm core}/M_{\rm clump}$) as a function of the mass growth sequence. 
Panel (b) presents the ratio of the individual core mass to the total core mass ($M_{\rm core}/M_{\rm total~core}$) versus the mass growth sequence. 
The data are taken from the ASHES project. 
The orange circles, blue squares, green triangles, and red crosses represent the most massive, second most massive, third most massive, and least massive cores, respectively. 
The solid lines in corresponding colors indicate the least-squares best fits, while the shaded regions represent the 95\% confidence intervals. 
The Spearman and Kendall correlation coefficients are indicated in the top-left corner of each panel.}
\label{fig1}
\end{figure}

(1) As shown in Figure \ref{fig1}(a), the core-to-clump mass ratios for all four categories (most massive, second, third, and least massive cores) increase monotonically with the total core-to-clump mass ratio (i.e., the mass growth sequence). 
This strong positive correlation is statistically confirmed by the high Spearman correlation coefficients ($\rho \ge 0.65$) and significant $p$-values ($p = 0.00$). 
Furthermore, considering their overlapping 95\% confidence intervals, the growth rates (slopes) for the three most massive cores are virtually identical and close to 1. 
However, the slope for the least massive cores is slightly different from those of the three most massive ones.

(2) Figure \ref{fig1}(b) shows that the mass fraction of the most massive core relative to the total core mass remains remarkably constant at $\sim 25\%$ (i.e., $10^{-0.6}$), despite a two-order-of-magnitude variation in the mass growth sequence. 
The mass fractions for the second and third most massive cores are similarly constant at $\sim 16\%$ and $\sim 10\%$, respectively. 
They are systematically lower, maintaining a consistent hierarchical order.
This flat trend is robustly confirmed by the extremely weak correlation coefficients ($\rho = 0.04$, $0.06$, and $-0.07$, respectively) and the correspondingly high $p$-values ($p = 0.81$, $0.70$, and $0.65$).  
These $p$-values are well above the 0.05 significance level, statistically confirming that there is no significant dependence on the mass growth sequence; instead, their mass fractions remain fixed at specific constant values. 
In contrast to the other groups, the least massive cores (red crosses) exhibit a much larger scatter and a notably lower $p$-value ($p = 0.15$), paired with a negative correlation coefficient ($\rho = -0.24$).
Although this scattered distribution does not constitute a statistically significant trend, it is visually distinct from the other groups, hinting at a possible divergence in the underlying physical properties of this least massive core population.

We note that using the mass ratio (e.g., $M_{\rm core}/M_{\rm clump}$) reduces several major systematic uncertainties in absolute mass estimates. By adopting the same distance, gas-to-dust mass ratio ($R_{\rm gd} = 100$), and dust temperature ($T_{\rm dust} = 10$--$20$~K) for the cores and their host clumps \citep{2023ApJ...950..148M}, these parameters cancel out when calculating the ratio. However, because the clump and core masses are derived from continuum emission at different wavelengths (0.87~mm and 1.3~mm, respectively), the dust opacity ($\kappa_{\nu}$) and the Planck function ($B_{\nu}(T)$) do not cancel out completely. Instead, the term $[\kappa_{0.87\rm mm} B_{0.87\rm mm}(T)] /[\kappa_{1.3\rm mm} B_{1.3\rm mm}(T)]$ acts as a constant scaling factor. In logarithmic space (i.e., $\log_{10}(M_{\rm core}/M_{\rm clump})$), this scaling factor simply adds a constant vertical shift of $\sim 0.5$~dex to the data points. As a result, this uniform shift preserves both the slopes of the best-fit lines and the correlation coefficients (such as the Spearman and Kendall $\tau$). Therefore, the evolutionary trends and mass fraction correlations presented in this work depend mainly on the observed flux ratios, making our physical conclusions highly robust against the uncertainties in the assumed parameters.

Furthermore, while missing flux is a known limitation of interferometric observations, its impact on our core mass estimates is expected to be minimal.
As a general rule, spatial filtering disproportionately affects nearby sources, which subtend larger angular sizes on the sky.
However, the dense cores identified in the ASHES survey are inherently compact, with physical radii ranging from $\sim 5 \times 10^{-3}$ to $5 \times 10^{-2}$~pc and a median diameter of $\sim 5360$~au \citep{2023ApJ...950..148M}.
Even for the nearest clumps in our sample, at a distance of $d = 2.4$~kpc, the maximum angular diameter of the largest cores is $\sim 8\farcs6$, and the median angular diameter is only $\sim 2\farcs2$.
Both values are well below the maximum recoverable scale of $\sim 19''$ provided by the combined 12~m Array and 7~m ACA \citep{2019ApJ...886..102S}.
In this regime, the fluxes of these compact cores are robustly recovered across the full distance range of our sample \citep{2026arXiv260712396M}.

\vspace{-5pt}
\section{Analysis and Discussion} 

We present a new pattern of environment-dependent core co-evolution: 
the progenitor seeds destined to form high-mass cores establish their mass dominance at an earlier mass growth sequence (Section~\ref{sec:massive core}); 
supply-limited accretion drives the synchronized co-evolution of all cores, scaling up their masses, and maintaining the early-established mass dominance and hierarchy (Section~\ref{sec:supply_limited}); 
a significant population of low-mass cores emerges at a later time (Section~\ref{sec: late form of low-mass core}).

\subsection{Early Mass Dominance of the Central Massive Cores}\label{sec:massive core}

The ASHES survey targets 70~$\mu$m dark clumps, which probe a very early stage of high-mass star formation. 
Although ASHES does not identify high-mass prestellar cores, with the most massive strictly prestellar core having a mass of only $\sim 10~M_{\odot}$, the most massive cores embedded in these young clumps, considering all core evolutionary classes, have already accumulated substantial masses, ranging from 1.7 to 81.1~$M_{\odot}$ with a mean of 12.5~$M_{\odot}$ \citep{2023ApJ...950..148M}. 
Because several of these most massive embedded cores already exceed the 27~$M_{\odot}$ threshold required to form a high-mass star, assuming a core-to-star formation efficiency of 30\% \citep{2019ApJ...886..102S,2023ApJ...950..148M}, their initial formation and mass accumulation must have occurred before the evolutionary window currently probed by the ASHES observations.

Our statistical analysis of the core mass distribution (see Figure \ref{fig1}) provides compelling evidence that the progenitor seeds of these central massive cores establish their mass dominance very early. As shown in Figure \ref{fig1}(a), the linear fits for all core mass lines exhibit a slope of approximately 1. This points to a synchronous and proportional growth mechanism in which both low- and high-mass cores grow simultaneously at rates proportional to their existing masses. 
Because the current accretion phase simply maintains the relative mass differences rather than creating them, the mass dominance of the massive core must have been established in an earlier epoch.

Furthermore, we observe that the most massive cores tend to possess a fixed fraction of the total dense gas contained within the cores. 
As revealed in Figure \ref{fig1}(b), the initial mass distribution is unequal. 
The $y$-intercept of the linear fit for the most massive cores (the orange line) is approximately $-0.6$, which corresponds to a mass fraction of $\sim$ 25\%. 
Given that a typical clump contains an average of 21 cores, the fact that a single massive core holds a quarter of the total core mass highlights a significant structural hierarchy. 
More importantly, the linear fit for the most massive cores in Figure \ref{fig1}(b) is flat (Spearman $\rho = 0.04$, $p = 0.81$), demonstrating that their mass fraction remains constant across the entire observed mass growth sequence.

If the progenitor seeds of the most massive cores had formed simultaneously with the low-mass population and competed for gas from the same starting point, one might expect an increasing trend in which their mass fraction gradually rises toward the currently observed value of $\sim 25\%$.
The absence of such an increasing trend suggests that the progenitor seeds of the most massive cores had already acquired a substantial mass advantage before the earliest stages captured in the ASHES observations.  
This early advantage, in our interpretation, reflects a distinct local dynamical environment.
In this environment, high density, converging inflows, and turbulence allow the central progenitor seeds to undergo super-Jeans fragmentation and rapidly become massive on core scales.
This physical picture is discussed in detail in Section~\ref{sec:super_jeans}.
From a physical standpoint, the early establishment of this mass dominance does not imply that the central massive cores are born fully formed. 
Like all cores, they inherently originate from initial low-mass seeds. 
The flat mass fraction highlights the critical role of distinct local physical conditions: the specific progenitor seeds destined to form high-mass stars acquire an extreme evolutionary head start at the cluster center.
This rapid early evolution is naturally explained by super-Jeans fragmentation, in which fast mass accumulation increases the effective fragmentation scale and allows the central seed to form directly as a larger-scale, more massive fragment (i.e., fragmenting less).

What our following analysis highlights is a sequential mass-growth picture for the overall cluster: while the central seeds rapidly grow into high-mass stars by fragmenting less through super-Jeans fragmentation (see Section \ref{sec:super_jeans}), the vast majority of the broader low-mass core population is relatively delayed by their lower gas densities and the absence of such non-stationary effect from super-Jeans fragmentation and forms continuously during the subsequent evolution (see Section \ref{sec: late form of low-mass core}). 
These coupled phenomena---reduced fragmentation and earlier mass dominance---are physically governed by the theoretical framework of non-stationary, transport-driven fragmentation. 
Note that the current angular resolution of this project is not high enough to resolve potential binary or multiple systems within these cores.

\vspace{-5pt}
\subsubsection{Super-Jeans Fragmentation} \label{sec:super_jeans}

If fragmentation is driven by thermal Jeans instability, the central thermal Jeans mass can be estimated from the averaged density ($n_{\rm c}$) and dust temperature ($T_{\rm d}$) in this region, 
\begin{equation}
M_{\text{th-J}} = \frac{c_s^3}{6} \left( \frac{\pi^5}{G^3 \rho} \right)^{1/2} = \frac{c_s^3}{6} \left( \frac{\pi^5}{G^3 \mu m_H n_c} \right)^{1/2} = 0.912 \left( \frac{T_{\text{d}}}{10 \, \text{K}} \right)^{3/2} \left( \frac{n_c}{10^5 \, \text{cm}^{-3}} \right)^{-1/2} \, M_{\odot}
\end{equation}
where $G$ is the gravitational constant (6.67 $\times$ 10$^{-8}$ cm$^3$ g$^{-1}$ s$^{-2}$) and $c_{\rm s}$ = [$k_{\rm B}$$T_{\rm d}$/($\mu$$m_{\rm H}$)]$^{1/2}$ is the speed of sound at temperature $T_{\rm d}$. 
The typical values for dense cores in IRDCs are $n_{\rm c} \sim 5 \times 10^{5}$~cm$^{-3}$ 
\citep{2018ARA&A..56...41M, 2019ApJ...886..102S} and $T_{\rm d} \sim 20$~K 
\citep{2019ApJ...886..102S, 2020ApJ...901..145F}, giving a thermal Jeans mass of $M_{\rm th-J} \sim 1.15$~$M_{\odot}$.
For the ASHES sample, the derived thermal Jeans masses range from 1.1 to 4.5~$M_{\odot}$, with a mean of 2.5~$M_{\odot}$ \citep{2019ApJ...886..102S}.
This thermal Jeans mass is almost an order of magnitude below the observed mass of the most massive core.

What we observe is that the most massive cores are associated with the highest-density regions, indicating a clear connection between massive cores and dense environments \citep[e.g.,][]{2019ApJ...886..102S,2023ApJ...950..148M,2024ApJS..270....9X}.
This difficulty can be explained by the Jeans fragmentation, where we expect the mass of the fragment to scale with the density of $m \sim \rho^{-1/2}$ in the IRDC (i.e., higher density yielding smaller fragments). 
It is likely that the initial mass of the fragment is systematically higher than the thermal Jeans mass from the start (i.e., it fragmented less). 
This is consistent with the transport-driven super-Jeans fragmentation scenario \citep{2024MNRAS.528.7333L}, in which super-Jeans is expected to occur at dynamically active sites---such as clump centers---where converging accretion flows drive a rapid increase in local density.
At such sites, externally driven converging flows can increase the background density on a mass-accumulation timescale comparable to or shorter than the local free-fall time, $t_{\rm acc}\lesssim t_{\rm ff}$. 
This condition does not imply gravitational collapse faster than free fall. 
Instead, the background density evolves substantially while local perturbations are developing.  
Under these non-stationary conditions, continuous mass injection favors larger fragmentation modes and increases the characteristic fragmentation length and mass ($\lambda_{\rm Jeans,\,acc} = \lambda_{\rm Jeans} \left(1 + t_{\rm ff}/t_{\rm acc}\right)^{1/3}$, $m_{\rm Jeans,\,acc} = m_{\rm Jeans} \left(1 + t_{\rm ff}/t_{\rm acc}\right)$) \citep{2024MNRAS.528.7333L}.
The fact that super-Jeans fragmentation can occur at the centers has been found for star-forming regions by \cite{10.1093/mnras/staf1116} using triangulation analysis, and our results suggest that transport-driven super-Jeans fragmentation may already operate during the earliest stages of clump evolution.

Fundamentally, this non-stationary transport-driven mechanism \citep{2024MNRAS.528.7333L} acts in concert with the ``delayed fragmentation'' scenario within the GHC framework \citep{2019MNRAS.490.3061V}.
The distinct evolutionary timescales between the central region and the surrounding envelope are driven by their contrasting local densities and non-stationary dynamical conditions.
At the central hub of a collapsing clump ($<0.1$~pc), converging inflows can create a dynamic, non-stationary environment.
Fed by the surrounding gas reservoir, these inflows can increase the background density on a mass-accumulation timescale comparable to or shorter than the local free-fall time ($t_{\rm acc}\lesssim t_{\rm ff}$).
This rapid, non-stationary mass accumulation can promote transport-driven super-Jeans fragmentation by increasing the characteristic fragmentation scale and mass, thereby suppressing the relative growth of small-scale fragmentation modes and favoring the formation of larger, more massive fragments. 
This process may enable the central progenitor seeds to establish their mass dominance earlier than the broader low-mass core population builds up in the surrounding, lower-density gas.
In contrast, the extended surrounding envelopes ($\sim 1$~pc) lack such non-stationary inflow acceleration and possess lower initial gas densities.
Because the characteristic timescale for gravitational collapse scales inversely with the square root of the density ($t_{\rm ff} \propto \rho^{-0.5}$), fragmentation in the outer regions is naturally much slower.
As a result, the formation of lower-mass cores in the outer regions is delayed. 
This contrast in density-regulated timescales, together with the differences in non-stationary dynamical conditions, naturally manifests as the ``delayed fragmentation'' scenario described in the GHC framework \citep{2019MNRAS.490.3061V}.
In this picture, the fragmentation of the surrounding gas is not instantaneous; instead, the outer gas must undergo global contraction sufficient to raise its local density by one to two orders of magnitude before localized Jeans fragmentation can efficiently proceed \citep{2019MNRAS.490.3061V}.
This required time delay prevents the immediate formation of smaller fragments, granting the central massive core a crucial head start. Consequently, the central core is able to accumulate a substantial amount of mass via continuous large-scale inflows well before the surrounding low-mass members can form and emerge.
As we will demonstrate quantitatively in Section~\ref{sec: late form of low-mass core} through analysis of the Gini coefficient and local free-fall times, the progressive build-up of the low-mass core population at later evolutionary stages provides observational evidence for this delayed fragmentation process in the ASHES sample.

This unified physical picture directly reconciles the two scenarios discussed earlier: the central massive core is inherently more massive and larger (because it fragmented less via super-Jeans fragmentation), and it established its mass dominance earlier (because converging inflows and higher local densities dictate that the center evolves fastest, while lower densities naturally delay the low-mass members). 
Thus, the massive core establishes its early mass dominance from an initial seed, matching the sequential evolutionary pattern observed in the ASHES sample.

\subsection{Supply-limited Accretion and Synchronized Co-evolution}\label{sec:supply_limited}

Along the mass growth sequence, Figure~\ref{fig1}(a) shows that the mass of the most massive core increases significantly, spanning almost two orders of magnitude across the sample.
This clear increase indicates that the central progenitor seed rapidly bypasses small-scale fragmentation to achieve a substantial mass very early on, and then continues to accrete steadily from a large gas reservoir. 
This steady accretion is sustained because the surrounding lower-density gas evolves on a much longer free-fall timescale, preventing its immediate fragmentation and allowing it to be continuously funneled inward to feed the central core.

As shown in Figure \ref{fig1}(a) and (b), the mass fractions of cores at all ranks---except the least massive---remain constant along the mass growth sequence, well within the 95\% confidence intervals. 
Consequently, the masses of all cores increase simultaneously by roughly the same factor. 
This demonstrates that cores of different masses grow synchronously during the accretion stage. 
Mathematically, maintaining a constant mass fraction ($M_{\rm core}/M_{\rm total~core} \approx {\rm constant}$) indicates that the mass accretion rate of each core is proportional to the total accretion rate of the system, i.e., $\dot{M}_{\rm core}/\dot{M}_{\rm total} \approx {\rm constant}$.
Physically, this corresponds to a coordinated accretion process, representing a scenario of mass-dependent accretion regulated by a limited reservoir, where cores partition the total infalling gas ($\dot{M}_{\rm total}$) proportionally to their current masses:
\begin{equation} \label{equation2}
    \dot{M}_i = \dot{M}_{\rm total} \frac{M_i}{\sum M_i},
\end{equation}
where $M_i$ and $\dot{M}_i$ represent the current mass and mass accretion rate of the $i$-th core, respectively, and $\sum M_i$ is the total mass of all cores within the clump.
The term $\dot{M}_{\rm total}$ denotes the overall gas infall rate supplied by the clump to the entire core system. 
In this framework, the accretion rate of each individual core ($\dot{M}_i$) is simply its mass-weighted share of the total gas supply.
This proportional growth ensures that the mass hierarchy established during the early fragmentation stage is preserved throughout the subsequent evolution: rather than altering the relative mass ranking or spacing among cores, the accretion process scales up the mass of each core synchronously.

\begin{figure}[!htbp]  
\centering
\includegraphics[width=0.99\textwidth]{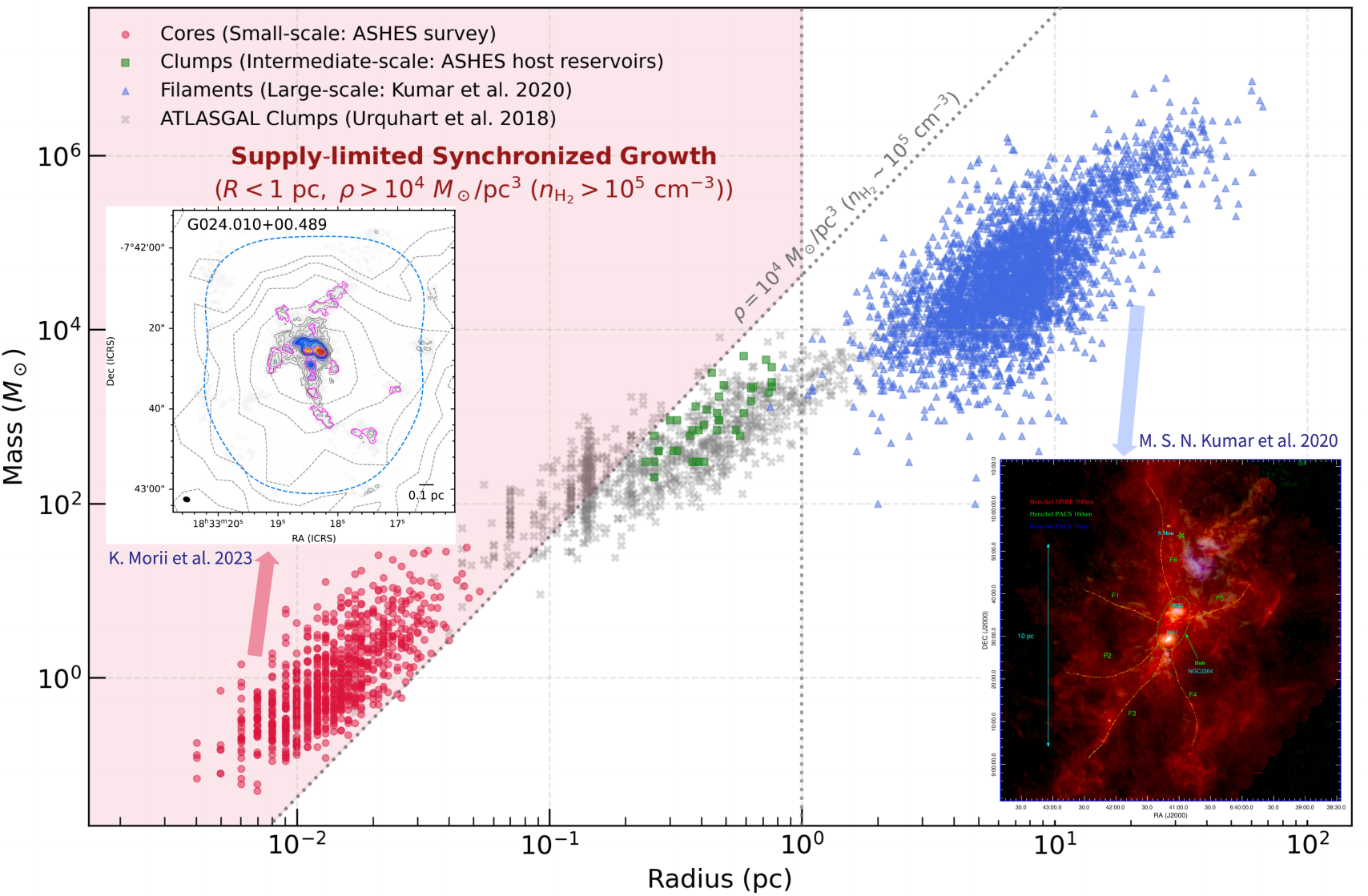}
\caption{
Mass--radius diagram of gas structures across distinct physical scales. 
Red circles indicate the small-scale dense cores ($\sim$ 0.1 pc), and green squares denote their intermediate-scale host clumps ($\sim$ 1 pc) that act as mass reservoirs, both identified in the ASHES survey \citep{2023ApJ...950..148M}.
Gray crosses represent the complete sample of Galactic massive clumps from the ATLASGAL survey \citep{2018MNRAS.473.1059U}. 
Blue triangles represent large-scale filaments ($\sim$ 10 pc) from \citet{2020A&A...642A..87K}. 
The inset images illustrate the real physical morphologies corresponding to the data: the bottom-right inset shows the large-scale hub-filament network (reproduced with permission from \citealt{2020A&A...642A..87K}, \copyright ESO), while the top-left inset shows the small-scale dense-core population resolved by ALMA in the ASHES survey \citep{2023ApJ...950..148M}.
The dashed diagonal and vertical lines mark the adopted boundaries of $\rho = 10^4\ M_\odot\ \mathrm{pc}^{-3}$ ($n_{\mathrm{H}_2} \sim 10^5\ \mathrm{cm}^{-3}$) and $R = 1\ \mathrm{pc}$, respectively.
The red shaded region ($R < 1\ \mathrm{pc}$, $\rho > 10^4\ M_\odot\ \mathrm{pc}^{-3}$) delineates the proposed applicability domain of the supply-limited synchronized-growth framework in this work, namely dense, sub-parsec core--clump systems.
}
\label{fig2:mass_radius}
\end{figure}

These results reveal a fundamental distinction between our findings and classical competitive accretion models \citep[e.g.,][]{2001MNRAS.323..785B,2001MNRAS.324..573B}. 
In particular, Bondi-Hoyle accretion models ($\dot{M} \propto M^2$) intrinsically rely on a uniform and unrestricted local gas supply when describing the formation of high-mass stars.  
In this framework, because ``gas quickly concentrates towards the deepest part of the potential well'' \citep[][]{2001MNRAS.323..785B}, the central massive cores are effectively supplied with a dense, unimpeded gas reservoir. 
Under such locally gas-rich conditions, a core that is initially more massive undergoes ``runaway'' mass growth, which would destroy the initial mass hierarchy and cause the relative mass fractions to diverge over time.

In contrast, the observed constant mass fractions suggest that the core system is governed by a \textit{supply-limited} accretion mechanism, which is fundamentally tied to the hierarchical nature of the star-forming environment.
As illustrated by the multi-scale mass--radius diagram in Figure~\ref{fig2:mass_radius}, the full sample of Galactic massive clumps from the ATLASGAL survey \citep{2018MNRAS.473.1059U} forms a continuous structural bridge connecting the extended large-scale filaments \citep{2020A&A...642A..87K} to the localized small-scale cores studied here \citep{2023ApJ...950..148M}.
This multi-scale separation delineates distinct physical regimes through which gas must flow in an unbroken hierarchical cascade before reaching the innermost regions.
At sub-parsec scales ($R < 1$~pc), cores embedded in gas with densities $\rho > 10^4\ M_\odot\ \mathrm{pc}^{-3}$ (corresponding to $n_{\mathrm{H}_2} > 10^5\ \mathrm{cm}^{-3}$) enter the supply-limited synchronized growth regime delineated by the red shaded region in Figure~\ref{fig2:mass_radius}.
The adopted density boundary is motivated by the characteristic $n_{\rm H_2}\sim10^5$--$10^6~\mathrm{cm}^{-3}$ densities of dense central regions hosting high-mass cores \citep[e.g.,][]{2018ARA&A..56...41M, 2025ARA&A..63....1B}.
Consequently, the total gas inflow rate ($\dot{M}_{\rm total}$) reaching the central hub is regulated by both the intermediate intermediate-scale clump reservoir and the large-scale filamentary system.
Because this global gas supply rate is limited by the upstream large-scale structures, individual small-scale cores cannot sustain the runaway accretion characteristic of idealized, unrestricted environments.
Instead, the restricted global gas flux enforces the coordinated accretion process described by Equation~(\ref{equation2}), compelling the cores to share the available mass flux and grow synchronously.

This physical picture naturally leads to a synchronized co-evolution scenario. 
Analogous to the co-evolution observed between supermassive black holes and their host galaxy bulges---where both components grow synchronously from the same finite cold gas reservoir \citep[e.g.,][]{1998AJ....115.2285M, 2005Natur.433..604D, 2013ARA&A..51..511K}---the high-mass and low-mass cores in our sample co-evolve from a shared, globally regulated clump reservoir. 
Consequently, the ultimate mass of a high-mass core is not the result of late-stage competitive accretion from an unrestricted gas pool. 
Rather, it is the result of its progenitor seed establishing its mass dominance very early by super-Jeans fragmentation, subsequently undergoing proportional mass growth through supply-limited accretion in synchronization with the rest of the cluster.

\vspace{3pt}
\subsection{Stratified Structure and the Continuous Build-up of Low-mass Cores} \label{sec: late form of low-mass core}

\setcounter{footnote}{0}

\begin{figure}[!htbp]  
\centering
\includegraphics[width=1\textwidth]{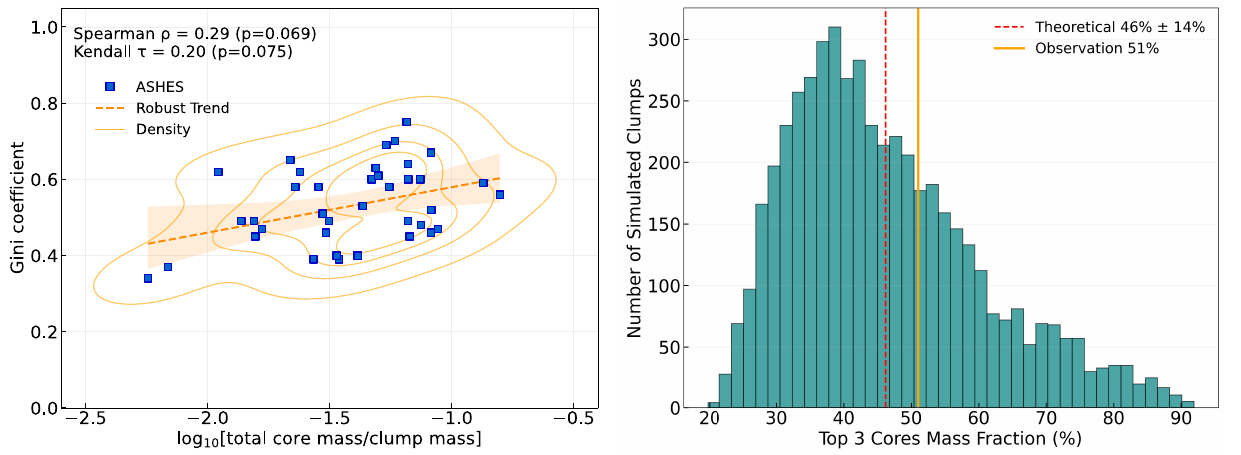}
\caption{Left panel: The relation between the Gini coefficient of the core-mass distribution and the mass growth sequence. 
The orange curve shows the density distribution, the orange dashed line represents the fitted trend, and the orange shaded area indicates the 95\% confidence interval of the fit. 
Spearman and Kendall correlation coefficients, along with different data markers, are provided in the top-left corner.
Right panel: The distribution of the mass fraction for the three most massive cores, derived from a Monte Carlo simulation. 
The red dashed line indicates the theoretical mean value ($46\% \pm 14\%$) from the simulation, while the solid orange line highlights our actual observed value ($51\%$).}
\label{fig3}
\end{figure}

Figure~\ref{fig3} presents the Gini coefficient\footnote{The Gini coefficient is a statistical measure of inequality or dispersion. Assuming non-negative values, its theoretical range is from 0 (perfect equality) to 1 (absolute inequality). While originally developed to measure wealth distribution, we adopt it here to quantify how uniformly the mass is partitioned among the cores within a single clump. Mathematically, for a clump containing $n$ cores with individual masses $M_1, M_2, ..., M_n$ and a mean core mass of $\bar{M}$, the Gini coefficient is calculated as half of the relative mean absolute difference: $G = \frac{\sum_{i=1}^n \sum_{j=1}^n |M_i - M_j|}{2 n^2 \bar{M}}$ \citep{10.1093/0198281935.001.0001}.} 
of the core mass distribution as a function of the mass growth sequence.
The Gini coefficient increases with the mass growth sequence, indicating the growing inequality of mass distribution.
Recalling our previous conclusion that the mass fractions of the most massive cores remain constant during this period (Figure \ref{fig1}b), this rising inequality cannot be driven by the massive cores monopolizing more mass.
Instead, it mathematically requires that the relative population of low-mass cores continuously increases over time at later evolutionary stages. 
These continuously forming low-mass cores individually add little to the total mass pool but significantly expand the lower-mass end of the population by number, thereby driving the overall distribution toward higher inhomogeneity.
With respect to the continued build-up of the low-mass core population, a related result is reported by \citet{2025A&A...696A.151C}. 
They find that the median maximum core mass among clumps increases with evolutionary stage, whereas the median minimum core mass remains approximately constant, and suggest that this persistent low-mass end may include newly formed fragments throughout the evolutionary sequence. 
These findings are qualitatively consistent with our interpretation that the low-mass core population continues to be replenished along the inferred mass growth sequence.
The growing contrast between the highest- and lowest-mass members is also compatible with the increasing mass inequality traced by the Gini coefficient.

In the ASHES sample, the three most massive cores account for $\sim 51\%$ of the total core mass. 
In a fully sampled standard Initial Mass Function (IMF), low-mass objects dominate the population by number and collectively contribute a substantial fraction of the total mass.  
This high mass fraction should therefore be interpreted in the context of the current ASHES detection limits and the early evolutionary stage of the targets.
To test whether the observed mass fraction of these top three cores is statistically unusual under the ASHES observational constraints, we performed a Monte Carlo simulation using the shape of the \citealt{2001MNRAS.322..231K} IMF as a reference mass spectrum.
Although the Kroupa IMF is defined for stellar masses, here we use its functional form only as a convenient reference distribution for a finite-sample truncation test, not as a physical model of the core mass function.
If a constant core-to-star formation efficiency is assumed, it corresponds to a uniform rescaling of the mass axis and therefore does not affect the relative mass fractions considered here. 
We adopted a lower core-mass limit of $M > 0.58\,M_{\odot}$ and a sample size of 21 cores per clump, matching the observational limits of \citet{2023ApJ...950..148M}.
As shown in the right panel of Figure~\ref{fig3}, under these ASHES-like truncation conditions, the three most massive objects account for $\sim 46\%$ of the total detectable mass, consistent with the observed value of $\sim 51\%$. 
This result indicates that the observed top-three mass fraction is not a statistical outlier within the currently detectable core population.
The high top-three mass fraction therefore reflects an underrepresentation of low-mass cores relative to the high-mass population at the current stage. 
Such an underrepresentation may arise either because some low-mass cores fall below the present detection limit or because a substantial fraction of the low-mass core population has not yet formed at these early stages.

The increasing Gini coefficient along the mass growth sequence, combined with the nearly constant mass fractions of the massive cores, supports the interpretation that the low-mass core population is not simply hidden from view but is continuously emerging at later evolutionary stages.
This continuous build-up of low-mass cores suggests that the Core Mass Function (CMF) is not necessarily static during the early stages of clump evolution.
Several studies have reported or proposed an evolutionary pathway in which an initially top-heavy CMF, meaning a shallower slope than the standard Salpeter IMF, subsequently steepens toward a more Salpeter-like distribution \citep{2019ApJ...886..102S,2024A&A...690A..33L}.
If an initially shallow CMF subsequently receives a sufficiently large contribution from continuously formed low-mass cores, its slope may steepen toward a more Salpeter-like distribution.
Our analysis, however, does not directly measure the evolution of the CMF slope. 
We therefore present this low-mass-core build-up as a possible pathway for the evolution of the initially shallow CMFs reported in some young regions, rather than as a universal CMF evolutionary sequence.
Actually, CMF evolution is likely more complex: an initially steep or Salpeter-like CMF may instead flatten and become increasingly top-heavy \citep{2023A&A...674A..75N,2023A&A...674A..76P,2025A&A...696A.151C}, and its evolution may depend on the interplay among mass-dependent core accretion, the continued formation of new cores, core subfragmentation and multiple-system formation, mass segregation, depletion of the clump-scale reservoir, and stellar feedback.

Importantly, this result does not imply that all low-mass cores form exclusively at late stages.
Rather, the emergence of the low-mass core population is a continuous and environment-dependent process.
While low-mass seeds certainly exist at early times---and may form even earlier and continuously over a prolonged timescale of $\sim 10^6$~yr within extended surrounding structures, as predicted by the large-scale hub-filament scenario \citep[e.g.,][]{2020A&A...642A..87K}---localized fragmentation in the immediate clump envelope is naturally delayed due to its lower gas density and longer free-fall time.
Thus, the continued supplementation of an already existing low-mass core population as the clump evolves can also contribute to the observed increase in the Gini coefficient.

This spatially segregated formation sequence and the delayed emergence of low-mass cores can be explained within a unified, environment-dependent fragmentation framework driven by both density contrasts and local dynamical non-stationary inflows within the clump.
From a density-driven perspective, the local free-fall time is determined by the local gas density as $t_{\rm ff} = [3\pi / (32G\rho)]^{1/2} = [3\pi / (32G\mu_{\rm H_2} m_{\rm p} n_{\rm H_2})]^{1/2}$, where $G$ is the gravitational constant, $m_{\rm p}$ is the proton mass, $n_{\rm H_2}$ is the H$_2$ number density, and $\mu_{\rm H_2} = 2.8$ is the mean molecular weight per hydrogen molecule.
In the dense central hubs where high-mass cores reside, the gas can reach number densities of the order of $10^5$--$10^6$~cm$^{-3}$ (e.g., \citealt{2018ARA&A..56...41M, 2025ARA&A..63....1B}). 
Adopting a representative intermediate value of $n_{\rm H_2} = 5 \times 10^5$~cm$^{-3}$ for our calculation, the estimated free-fall time is $t_{\rm ff} \sim 4 \times 10^4$~yr. 
In contrast, the more diffuse surrounding structures where low-mass cores predominantly reside have a density threshold on the order of $n_{\rm H_2} \sim 10^4$~cm$^{-3}$ (e.g., \citealt{2010ApJ...724..687L, 2014ApJ...782..114E}); adopting this representative value as the density for low-mass regions, the free-fall time is $t_{\rm ff} \sim 3 \times 10^5$~yr.
This baseline difference in evolutionary timescales is further amplified by the local dynamical environments.
In the dense central regions, converging accretion flows can create a non-stationary environment, causing the background density to increase on a mass-accumulation timescale comparable to or shorter than the local free-fall time. 
Under these conditions, continuous mass injection increases the characteristic fragmentation scale and mass, bypassing small-scale fragmentation, through transport-driven super-Jeans fragmentation \citep{2024MNRAS.528.7333L, 10.1093/mnras/staf1116} and enabling the central massive-core progenitors to establish an early mass advantage relative to the surrounding low-mass core population.
Conversely, the lower-density surrounding regions lack this non-stationary acceleration. Fragmentation here is governed primarily by the slower, density-driven Jeans collapse.
The combination of lower gas densities (longer $t_{\rm ff}$) and the absence of rapid dynamical pile-up therefore acts to delay fragmentation in the surrounding envelopes. This environment-dependent mechanism naturally produces a scenario in which massive cores establish their mass dominance rapidly at the clump center, while the low-mass core population forms through a delayed fragmentation process that builds up continuously over time.

It is important to place these localized, environment-dependent findings within the broader large-scale context of massive cloud evolution.
Star formation is globally driven by filamentary networks that govern large-scale gas assembly \citep[e.g.,][]{2009ApJ...700.1609M, 2020A&A...642A..87K}.
In the large-scale hub-filament picture, low-mass stars assemble early and continuously within the extended, parsec-scale filamentary arms ($\sim 10$--$20$~pc), forming slowly over timescales of $\sim 10^6$~yr \citep{2020A&A...642A..87K}. 
At these large distances, the gas is unaffected by the non-stationary dynamics of the central region, allowing local Jeans fragmentation to proceed over longer timescales.
Longitudinal gas flows channel material along these filaments into the central junction (the hub), creating a dense, amplified environment with a massive accretion reservoir.

However, the ASHES observations do not probe these extended large-scale filaments; instead, they resolve the compact environments within massive clumps on scales of $\lesssim 1$~pc.
Within this localized region, the physical conditions are governed by the deep gravitational potential of the central mass concentration.
Because the gas density and the intensity of converging inflows scale with radial distance toward the center, the immediate clump envelope ($\sim 0.1$--$1$~pc) represents a transitional zone. 
Here, the gas density is lower, and the inflows lack the non-stationary geometric convergence found at the very center, which naturally acts to delay the formation of low-mass cores in the immediate vicinity of the central hub.
Meanwhile, the innermost hub ($\lesssim 0.1$~pc) is where the converging flows reach their maximum intensity, creating a non-stationary environment that triggers rapid super-Jeans fragmentation. 
Within the hub-filament systems paradigm, the density-amplified hub and its replenishing filamentary reservoir sustain high accretion rates, allowing these seeds to grow rapidly into high-mass stars on a characteristic formation timescale of $\sim10^5$~yr \citep{2020A&A...642A..87K}.
This behavior is qualitatively consistent with the transport-driven super-Jeans framework that the progenitor seeds destined to become massive cores establish their mass dominance at a very early stage across the non-stationary acceleration.

With these spatial scales clearly defined, a consistent picture emerges: while low-mass star formation may begin at early times in the distant large-scale filaments ($\sim 10$~pc), the formation of low-mass cores in the immediate environment of the clump ($\sim 1$~pc) is naturally delayed by their lower local gas densities and the absence of non-stationary inflow acceleration.
The proposed environment-dependent fragmentation scenario on small scales therefore fits within the large-scale hub-filament framework, explaining why the massive core establishes its mass dominance first at the clump center while the surrounding low-mass core population builds up continuously at later evolutionary stages.

\section{Conclusion} 

Using 839 cores from 39 High-mass clumps at early stages in the ASHES project, and studying the co-evolution of cores with different masses embedded in the same clump across mass growth sequences.

Our findings can be summarized as follows:
\begin{itemize}
    \item The progenitor seeds destined to become massive cores establish their mass dominance at a very early stage, where the clump often exhibits a limited degree of fragmentation.
    \item The mass of the most massive core is larger than the thermal Jeans mass, indicating that super-Jeans fragmentation plays a key role in the fragmentation related to the formation of the most massive cores.
    \item The mass growth mode of the cores is self-similar, synchronous, and proportional, reflecting a scenario of mass-dependent accretion with reservoir regulation [$\dot{M}_i = \dot{M}_{\rm total} (M_i / \sum M_i)$]. Through this mechanism of synchronized growth, the progenitor seeds of massive cores are able to maintain the initial mass dominance and hierarchy they established during the very early fragmentation stage. 
    \item The system follows an environment-dependent fragmentation mechanism driven by spatial stratification: rapid non-stationary mass accumulation in the dense central regions favors transport-driven super-Jeans fragmentation, enabling central progenitor seeds to establish an early mass advantage, while lower gas densities and the absence of non-stationary inflow acceleration regulate and delay the continuous build-up of the surrounding low-mass core population.
    \item The relative contribution of the low-mass cores population continuously increases over time, driving up the overall degree of fragmentation.
\end{itemize}

In summary, our results reveal an environment-dependent fragmentation scenario: central progenitor seeds rapidly evolve into massive cores via super-Jeans fragmentation and maintain their mass dominance through synchronized, supply-limited accretion, while low-mass cores are relatively delayed due to their lower gas densities and the lack of non-stationary acceleration, and continuously emerge at later stages. 
This synchronized, supply-limited growth regime operates specifically at sub-pc scales ($R<1$~pc) with gas densities $n_{\rm H_2} > 10^5~{\rm cm}^{-3}$, where global reservoir regulation governs the accretion process.

\begin{acknowledgments}
GXL acknowledges support from NSFC grant No.12273032.
PS was partially supported by a Grant-in-Aid for Scientific Research (KAKENHI Number JP26H02066 and JP23H01221) of JSPS.
\end{acknowledgments}

\bibliography{sample7}{}
\bibliographystyle{aasjournalv7}

\vspace{12pt}
\noindent\textbf{ORCID:}\\
Dan Miao: \url{https://orcid.org/0000-0001-9798-9852}\\
Guang-Xing Li: \url{https://orcid.org/0000-0003-3144-1952}\\
Patricio Sanhueza: \url{https://orcid.org/0000-0002-7125-7685}

\end{document}